\documentclass[preprint2]{aastex61}
\usepackage{amsmath}
\usepackage{natbib}
\usepackage{hyperref}
\hypersetup{
  colorlinks, linkcolor=blue
}
\begin{document}

\bibliographystyle{apj}

\title{An Observational Diagnostic for Distinguishing Between Clouds and Haze in Hot Exoplanet Atmospheres}

\author{Eliza M.-R. Kempton}

\affil{Department of Physics, Grinnell College, 1116 8th Ave., Grinnell, IA 50112, USA}

\email{kemptone@grinnell.edu}

\author{Jacob L. Bean}

\affil{Department of Astronomy \& Astrophysics, University of Chicago, 5640 S. Ellis Avenue, Chicago, IL 60637, USA}

\author{Vivien Parmentier}

\altaffiliation{NASA Sagan Fellow}

\affil{Department of Planetary Sciences and Lunar and Planetary Laboratory, The University of Arizona, Tucson, AZ 85721, USA}

\begin{abstract}

The nature of aerosols in hot exoplanet atmospheres is one of the primary vexing questions facing the exoplanet field. The complex chemistry, multiple formation pathways, and lack of easily identifiable spectral features associated with aerosols make it especially challenging to constrain their key properties. We propose a transmission spectroscopy technique to identify the primary aerosol formation mechanism  for the most highly irradiated hot Jupiters (HIHJs).  The technique is based on the expectation that the two key types of aerosols -- photochemically generated hazes and equilibrium condensate clouds -- are expected to form and persist in different regions of a highly irradiated planet's atmosphere.  Haze can only be produced on the permanent daysides of tidally-locked hot Jupiters, and will be carried downwind by atmospheric dynamics to the evening terminator (seen as the trailing limb during transit).  Clouds can only form in cooler regions on the night side and morning terminator of HIHJs (seen as the leading limb during transit).  Because opposite limbs are expected to be impacted by different types of aerosols, ingress and egress spectra, which primarily probe opposing sides of the planet, will reveal the dominant aerosol formation mechanism.  We show that the benchmark HIHJ, WASP-121b, has a transmission spectrum consistent with partial aerosol coverage and that ingress-egress spectroscopy would constrain the location and formation mechanism of those aerosols.  In general, using this diagnostic we find that observations with \emph{JWST} and potentially with \emph{HST} should be able to distinguish between clouds and haze for currently known HIHJs.

\end{abstract}


\section{Introduction \label{intro}}

The prevalence of aerosols in hot Jupiter atmospheres has been definitively shown through transmission spectroscopy \citep{sin16}. Yet the nature of these aerosols remains unconstrained and is a primary challenge in the modeling and interpretation of exoplanet spectra. Two distinct mechanisms have been proposed for the formation of hot Jupiter aerosols. (1) Clouds are formed via direct condensation of atmospheric gases, and (2) hazes are formed via chemical reaction pathways in the atmosphere that typically begin with UV photolysis\footnote{http://www.planetary.org/blogs/guest-blogs/2016/0324-clouds-and-haze-and-dust-oh-my.html}. Both mechanisms result in the formation of solid or liquid aerosols and produce the same qualitative observational consequences -- weaker than expected absorption features and strong scattering signatures in transmission spectra \citep[e.g.][]{kre14, sin16}. In principle the composition of aerosols can be determined via spectroscopy \citep[e.g.][]{mor15, gao17, wak17, rob17}, but in practice the signatures are often weakly wavelength dependent, degenerate with microphysical properties such as particle size and distribution, and/or are only detectable with observational facilities that will not be constructed for at least a decade.

The most highly irradiated hot Jupiters (HIHJs) provide an ideal laboratory to break the degeneracy between the two aerosol formation mechanisms. These planets, with equilibrium temperatures greater than approximately 2,000~K, have daysides that are too hot for condensation to occur. Consequently, cloud formation can take place only on the night side of a HIHJ or downwind of the anti-stellar point on the cooler morning terminator. Conversely, haze-forming photochemistry can only occur on a tidally-locked planet's dayside where the stellar UV flux precipitates the photolysis reaction pathway.  Therefore, rather than searching for the spectral signatures of distinct aerosol species, which can be challenging to uniquely identify, we outline a method to identify the aerosol formation pathway by constraining the \textit{location} on a HIHJ where aerosols are being generated.

\begin{figure}
\begin{center}
\includegraphics[scale = 0.42]{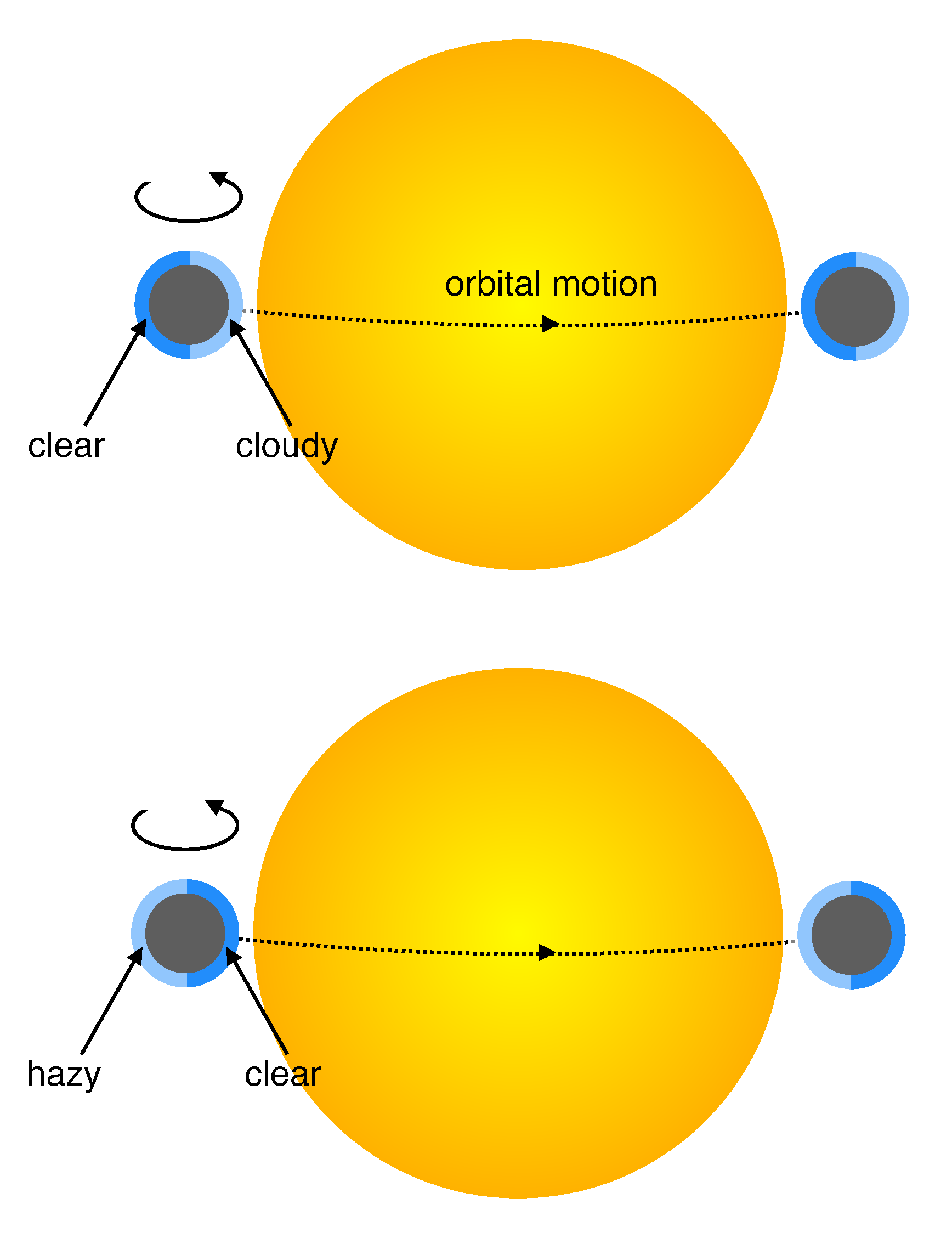}
\end{center}
\caption{Toy model for the distribution of clouds and haze over the terminator of a HIHJ. We assume that clouds can form on the night side and leading limb, whereas haze can be efficiently transported from the dayside to the trailing limb. This results in a divided terminator where one half is polluted by aerosols while the other half retains clear skies. Clouds will be seen preferentially in the ingress transmission spectrum, which primarily weights the leading limb, while hazes will be seen preferentially in the egress transmission spectrum.  \label{diagram}
}
\end{figure}

The proposed technique is grounded in the theory of atmospheric circulation for tidally-locked planets.  A universal prediction of hot Jupiter general circulation models (GCMs) is the occurrence of a strong band of equatorial winds moving in the direction of the planet's rotation at a pressure of $\sim$1 bar \citep[e.g.][]{sho09, rau10, dob10, hen11, kat16}.  The co-rotational winds are associated with the hotspot offsets seen in hot Jupiter phase curves \citep[e.g.][]{knu07, zel14, ste14}, because the hottest location of the planet is transported downwind of the substellar point.  This picture of atmospheric dynamics for hot Jupiters has two key consequences for aerosol formation. (1) Co-rotational winds will transport hazes produced on the dayside to the evening terminator (viewed as the planet's trailing limb during transit). (2) As previously mentioned, clouds will form readily on the cooler morning terminator (seen as the leading limb during transit) but not on the hotter evening terminator \citep{par16}. The observational consequence of this model (depicted in Figure~\ref{diagram}) is that the ingress transmission spectrum, which primarily probes the leading limb, is a diagnostic of clouds, while the egress spectrum diagnoses haze. By differencing the two, the side of the planet that is primarily dominated by aerosols will be revealed, thus constraining the dominant aerosol formation mechanism.

In distinguishing between aerosol formation mechanisms, we must first ask the question of whether one formation pathway is expected over the other. Possible evidence of night-side and/or morning terminator clouds has already been reported in hot Jupiter phase curves obtained in both the optical \citep{demo13, est15, hu15, shp15} and IR \citep{ste14, ste17, kat15}.  Chemical reaction networks leading to haze formation are challenging to compile, even for Solar System objects, but the pathways proposed in the literature to date are most efficient for cooler planets \citep{zah09, mos14, zah16}.  This combination of evidence points to night-side clouds as the more likely aerosol production pathway for HIHJs.  However, given uncertainties in haze formation models, \emph{spectral} confirmation of the location of aerosols would provide a strong constraint on the dominant formation mechanism.  In this work, we take an agnostic approach to the aerosol formation mechanism with the goal of letting the observational data make the final distinction.

The effects of partial aerosol coverage on exoplanet transmission spectra have previously been explored by \citet{lin16}, who found that patchy clouds can mimic high mean molecular weight cloud-free atmospheres. Additionally, both \citet{lin16} and \citet{von16} studied the effects of single-hemisphere aerosols on the shape of transit light curves during ingress and egress. In what follows, we extend the models of partial cloudiness to calculate the ingress and egress spectra of HIHJs, as a means to determine the aerosol formation mechanism. In Section~\ref{Targets}, we present ideal observational targets for ingress/egress spectroscopy, focusing on the test case of WASP-121b.  In Section~\ref{Model}, we describe our model for hazy and cloudy ingress and egress spectra. We present theoretical HIHJ spectra in Section~\ref{Results}. In Section~\ref{Conclude} we conclude by discussing additional complexities associated with our proposed aerosol formation diagnostic and commenting on the observability of ingress and egress spectra with \emph{HST} and \emph{JWST}.

\section{Target Selection \label{Targets}}

\begin{deluxetable*}{lcccccc}
\tabletypesize{\scriptsize}
\tablecolumns{7}
\tablewidth{0pc}
\tablecaption{Potential targets for cloud-haze discrimination}
\tablehead{
\colhead{Planet} &
\colhead{Mass (M$_{Jup}$)} &
 \colhead{Radius (R$_{Jup}$)} &
 \colhead{T$_{eq}$\tablenotemark{a} (K)} &
 \colhead{$m_{K}$} &
 \colhead{Transmission\tablenotemark{b}} &
 \colhead{Signal-To-Noise\tablenotemark{b}}
}
\startdata
WASP-76b &  0.92 & 1.83 & 2182 &  8.24 & 0.80 & 1.34\\
WASP-33b &  2.16 & 1.68 & 2735 &  7.47 & 0.43 & 1.04\\
WASP-121b &  1.18 & 1.87 & 2358 &  9.37 & 1.00 & 1.00\\
KELT-7b &  1.28 & 1.53 & 2050 &  7.54 & 0.32 & 0.73\\
KELT-9b &  2.88 & 1.89 & 4051 &  7.48 & 0.28 & 0.67\\
WASP-12b &  1.47 & 1.90 & 2562 & 10.19 & 0.72 & 0.49\\
WASP-19b &  1.14 & 1.41 & 2078 & 10.48 & 0.81 & 0.49\\
KELT-17b &  1.31 & 1.52 & 2087 &  8.65 & 0.34 & 0.48\\
WASP-82b &  1.25 & 1.71 & 2207 &  8.76 & 0.30 & 0.39\\
WASP-78b &  0.86 & 2.06 & 2354 & 11.01 & 0.71 & 0.33\\
\enddata
\tablenotetext{a}{Calculated assuming planet-wide redistribution}
\tablenotetext{b}{Normalized to WASP-121b}
\label{table1}
\end{deluxetable*}

The ten HIHJs that are expected to produce the highest signal-to-noise transmission spectra are listed in Table~\ref{table1}. The planet and host star parameters are taken from the TEPCat catalog for transiting exoplanets\footnote{http://www.astro.keele.ac.uk/jkt/tepcat/}. The size of the expected transmission signal scales proportionately to $R_{pl} H / R_{*}^{2}$, where $R_{pl}$ and $R_{*}$ are the planetary and stellar radii, respectively.  The transmission signal is calculated for spectral features corresponding to one atmospheric scale height, assuming solar composition and an atmospheric temperature equivalent to the planet's equilibrium temperature. The expected signal-to-noise is calculated per unit of exposure time using the relative brightnesses of the host stars in the K band. 

The planets in Table~\ref{table1} represent the best targets for high precision ingress and egress spectra to discriminate between clouds and haze. Of these planets, we focus the remainder of this Letter on more detailed modeling of the benchmark HIHJ WASP-121b. This planet has been shown to have water features in its transmission spectrum, along with strong hints of TiO, VO, and possibly FeH absorption \citep{eva16}. The only two targets expected to produce higher signal-to-noise spectra are WASP-76b and WASP-33b.  An initial study of the former planet does not suggest evidence for aerosols in its transmission spectrum \citep{tsi17}.  The latter has not yet been characterized in transmission and is also not an ideal target because its host star is a $\delta$-Scuti variable \citep{her11}.

\section{Model Description \label{Model}}

To determine the locations where cloud condensation will occur in WASP-121b, we use the SPARC/MITgcm global circulation model~\citep{sho09} to model its three dimensional atmospheric circulation and thermal structure. Our setup is similar to the one used in~\citet{par16}, but with planetary and system parameters chosen to match WASP-121b. Following the detection of TiO/VO in the transit spectrum of~\citet{eva16}, the model includes TiO/VO in solar composition as a gaseous compound, leading to the presence of a large thermal inversion on the dayside of the planet~\citep{hub03,for08,par15}. As shown by~\citet{par16}, the spatial distribution of clouds on a planet is only marginally affected by their radiative feedback on the atmospheric circulation. For simplicity, the radiative feedback of the cloud was therefore not considered in this simulation. The model was run for 300 days and all quantities were averaged over the last 100 days. 

Transmission spectra are generated using the \texttt{Exo-Transmit} radiative transfer package \citep{kem17} with the included opacity data \citep{fre08, fre14, lup14}, a solar composition atmosphere, and a 1-D double-gray T-P profile \citep{mil09} with an upper atmosphere isothermal temperature of 2100 K, which is consistent with a full planet average of the GCM output (see Section~\ref{Results}). Pairs of transmission spectra are generated for each planet -- one clear of aerosols and a second that has an added gray or Rayleigh scattering opacity to simulate the presence of clouds and/or haze. Note that the wavelength-dependence of the aerosol opacity (gray vs.\ Rayleigh scattering) is independent of the aerosol formation mechanism. Both clouds and haze are capable of possessing a wide range of spectral properties, which depend on the composition, size, and distribution of the particles.

Based on the toy model shown in Figure~\ref{diagram}, integrated ingress and egress spectra, $S$, are calculated by a weighted mean of 20 snapshots equally spaced in time as the planet passes over each side of the stellar limb:

\begin{equation}
S = \frac{1}{20}\sum_{n=1}^{20} (f_{clear} S_{clear} + f_{aer} S_{aer}), \label{eq1}
\end{equation} 

\noindent where $f_{clear}$ and $S_{clear}$ are the fraction of the clear-atmosphere side of the planet in front of the star and the clear transmission spectrum, respectively.  Similarly $f_{aer}$ and $S_{aer}$ describe the aerosol-dominated side of the planet.  For the example of a planet with morning-terminator clouds $f_{clear} = 0$ for the first half of ingress, and $f_{clear} = 1$ for the first half of egress.  The opposite is true for $f_{aer}$ during the second half of ingress and egress. Equation~\ref{eq1} assumes that the planet progresses linearly in front of the star and does not account for limb darkening. Wavelength dependent limb darkening will impact both the ingress and egress spectrum in the same manner, so the differential comparison between the two sides of the transit required by our aerosol diagnostic should also remain unaffected.

\section{Results  \label{Results}}

\begin{figure}
\includegraphics[scale = 0.47]{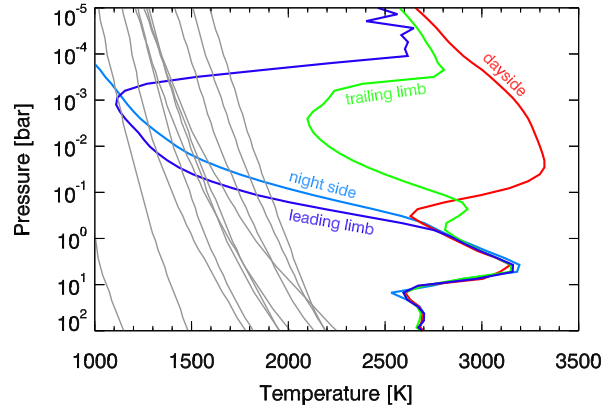}
\caption{T-P profiles for WASP-121b compared to condensation curves. The colored lined show equatorial T-P profiles from the WASP-121b GCM taken at the substellar point (dayside), anti-stellar point (night side), and the two limbs, as indicated.  The grey lines are condensation curves for known condensible species from \citet{mba16}, assuming solar composition.  Intersections between the T-P profiles and condensation curves indicate locations where clouds should form in the planet's atmosphere.   \label{t_p_fig}
}
\end{figure}

\begin{figure*}
\begin{center}
\includegraphics[angle=90, scale=0.63]{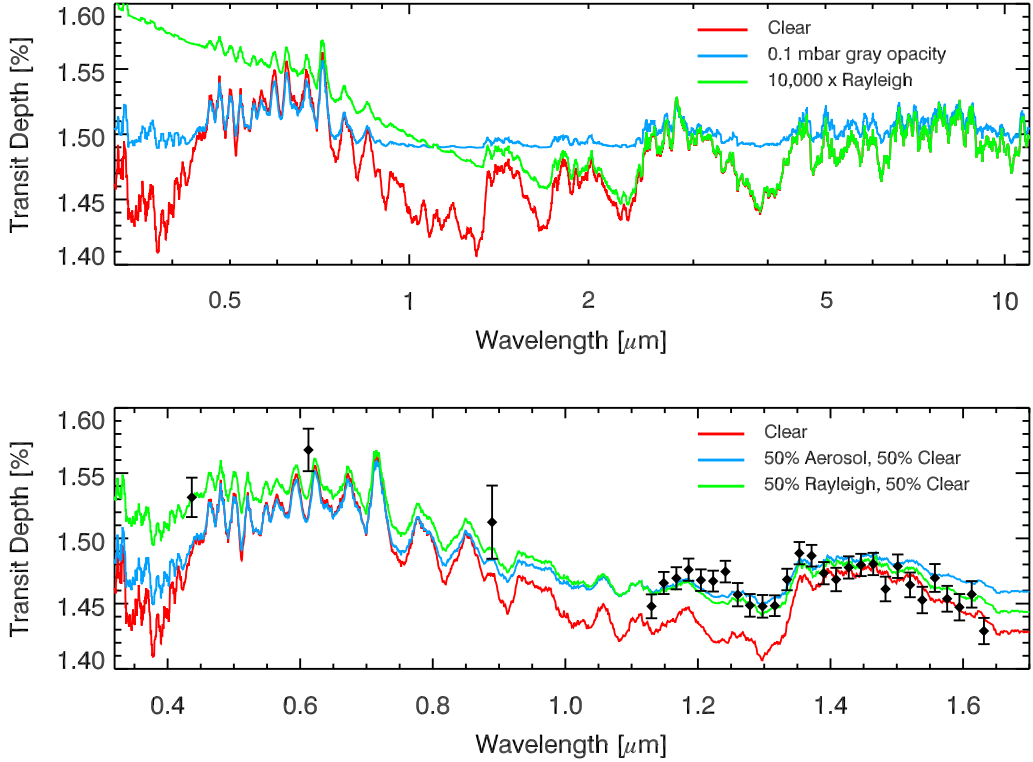}
\end{center}
\caption{Top: Transmission spectra for WASP-121b at solar composition.  The three models represent an aerosol-free atmosphere (red), an atmosphere with an optically thick gray opacity source at 0.1 mbar (blue), and an atmosphere with Rayleigh scattering of 10,000 times the nominal solar value (green). The latter two represent atmospheres with (uniform) aerosols.  Bottom: Models for single-hemisphere aerosol coverage of WASP-121b.  A 50\% weighting of the gray and clear spectra is shown in blue, and a 50\% weighting of the Rayleigh scattering and clear spectra is shown in green. The clear atmosphere spectrum is again shown in red, for reference.  Transmission spectrum data from \citet{eva16} (ground-based shortward of 1 $\mu$m and \emph{HST} WFC3 data longward of 1 $\mu$m) are overlaid in black showing good agreement between the existing data and the models with partial aerosol coverage.  \label{transmission_fig}
}
\end{figure*}

The resulting T-P profiles from our GCM of WASP-121b are shown in Figure~\ref{t_p_fig}.  Our qualitative predictions from Section~\ref{intro} are borne out in this detailed model.  That is, both the dayside and evening terminator of the planet remain too hot at all pressures for condensation of any species to occur.  However, the night side and morning terminator fall substantially below the condensation temperatures of a variety of condensibles across the pressure range probed by transmission spectroscopy ($\sim$1 mbar), which should result in morning-terminator clouds.

\begin{figure*}
\begin{center}
\includegraphics[angle=90, scale=0.6]{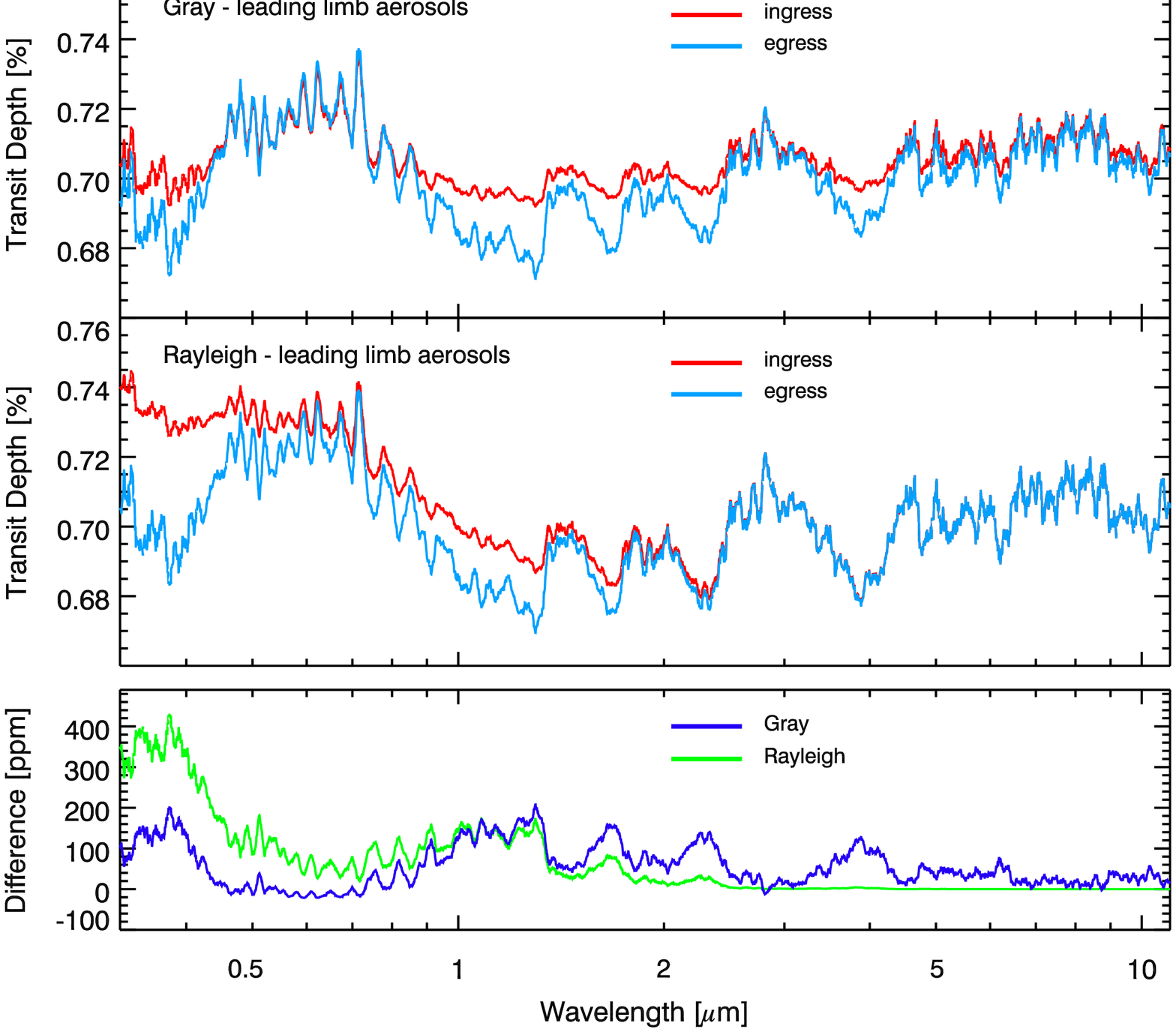}
\end{center}
\caption{Top: Ingress and egress spectra for partial aerosol coverage of WASP-121b. The planet is modeled to have single hemisphere aerosol coverage, with an optically thick gray aerosol opacity at 0.1 mbar (top panel) or with enhanced Rayleigh scattering at 10,000 times the nominal solar value (middle panel). If the planet has morning-terminator clouds, the ingress and egress spectra are shown in red and blue, respectively. If the planet has evening-terminator haze, the lines are reversed, and the ingress and egress spectra are shown in blue and red, respectively. Bottom: The difference between the ingress and egress spectra for the cases of gray aerosols (purple) and Rayleigh scattering aerosols (green). This is the observational diagnostic for single-hemisphere aerosols.   \label{diff_fig}
}
\end{figure*}

We generate three model transmission spectra (each assuming uniform terminator composition) for WASP-121b, shown in the top panel of Figure~\ref{transmission_fig} -- one clear solar composition atmosphere and two aerosol-dominated spectra. The latter are calculated in two limits of the aerosol opacity -- purely gray and purely Rayleigh scattering.  In the lower panel of Figure~\ref{transmission_fig}, the clear and aerosol-dominated spectra are averaged together with equal weighting to show the effects of single hemisphere aerosols on the center-of-transit transmission spectrum.  The gray aerosol opacity is selected to become optically thick at a pressure of 0.1 mbar, and the Rayleigh scattering opacity is set at 10,000 times the nominal solar value. These values were chosen to fit the observed transmission spectrum of WASP-121b, revealing that partial aerosol coverage and a solar composition naturally explain the existing data for this planet.  Note that the single-hemisphere aerosol models (the blue and green lines from the lower panel of Figure~\ref{transmission_fig}) do not require invoking FeH, which is an opacity source that is not currently included with the \texttt{Exo-Transmit} package, to explain the observed spectrum.

Integrated ingress and egress spectra are calculated according to Equation~\ref{eq1} and shown in Figure~\ref{diff_fig} for the limits of gray and Rayleigh scattering aerosols. In each case, one side of the transit (either ingress or egress) more clearly shows the spectral signature of aerosols, as expected. The aerosol-dominated side is reversed depending on whether the planet has morning-terminator clouds or evening-terminator haze. If the WASP-121b aerosols are primarily gray, the differences between the ingress and egress spectra appear most strongly in between absorption bands (around 1.3, 1.7, 2.3, and 3.9 $\mu$m) where the aerosol absorption causes the atmosphere to become optically thick at higher altitude. For a Rayleigh scattering aerosol, the spectral differences are also seen between absorption bands, with the strongest differences occurring at short wavelengths where the scattering signature is the strongest. In both cases, the typical magnitude of spectral differences between ingress and egress spectra for WASP-121b is $\sim$100 ppm, and as large as 400 ppm in the optical for a strongly Rayleigh scattering aerosol.

\section{Discussion and Conclusions  \label{Conclude}}

The detailed spectral models for WASP-121b shown in Figures~\ref{transmission_fig} and \ref{diff_fig} are just one realized example of how ingress and egress spectra can be used to constrain the aerosol formation mechanism for HIHJs.  In general, either the ingress or the egress spectrum should present weaker features depending on which formation mechanism is at work. The advantage of focusing on HIHJs is that haze and clouds should form and persist in distinctly different regions of the planet, thus providing a straightforward technique for constraining the primary pathway to aerosol formation on these planets. The techniques described in this Letter are also applicable to determining the global distribution of aerosols on hot Jupiters with equilibrium temperatures less than 2,000 K, but if dayside and evening-terminator aerosols are discovered the interpretation will be more ambiguous.

A third possible observational outcome is the case in which both sides of the planet are found to be equally affected by aerosols.  We argue that in HIHJs this outcome is also indicative of dayside haze.  Haze particles are far more robust than clouds.  The latter will evaporate on $\sim$1-min timescales upon coming into contact with regions of the atmosphere with temperatures above the condensation threshold (Powell et al., in prep.).  The former will persist until they are destroyed by chemical processes or settle out of the atmosphere.  Hazes that persist for longer than the $\sim$24-hr timescale for the super-rotating jet to cross the night-side hemisphere will therefore be present on both the morning and evening terminators.  Additionally, day-to-night winds predicted across the limb at pressures $\lesssim 1$ mbar or counter-rotating jets at mid-latitudes \citep[e.g.][]{mil12, kem14} could lead to high-altitude dayside haze being transported equally across both sides of the terminator.  While these two scenarios for uniform terminator haze coverage are plausible, it is challenging to come up with an equally plausible scenario resulting in clouds on both limbs for a HIHJ.

Our toy model from Figure~\ref{diagram} simplifies the atmospheric circulation patterns predicted by hot Jupiter GCMs in focusing on only the key feature of the co-rotating jet. More realistically, the morphology of the equatorial jet may not lead to a perfect hemispherical separation of aerosol vs.~non-aerosol dominated segments of the terminator.  For example, magnetic effects, which are expected to influence the most highly irradiated planets and are typically not self-consistently included in GCMs, could alter the magnitude and morphology of atmospheric winds \citep{rau13, rog14}. Additionally, GCMs that include cloud tracer particles qualitatively predict a non-uniform spatial distribution of aerosols in exoplanet atmospheres \citep{par13, cha15}. These non-uniform variations in aerosol coverage could be measured by looking at the change in the spectrum during the ingress and the egress ("transit mapping"), a technique that would require a much higher signal-to-noise that the integrated ingress/egress spectra proposed here.  Ultimately, fully self-consistent models that incorporate the chemistry, transport, and radiative effects associated with aerosols are not yet readily available for hot Jupiters and will require accurate input parameters that can only be constrained through observational studies, such as the ones proposed here.  

Temperature-related scale height differences between the leading and trailing terminators will also affect the relative magnitude of features in ingress and egress spectra, as noted previously by \citet{for10} and \citet{bur10}. The observational techniques developed in this Letter rely on disentangling the effects of scale height from those of aerosols, yet both morning-terminator clouds and cooler morning-terminator temperatures will diminish the strength of features in the ingress spectrum relative to egress.  To quantify the impact on the transmission spectrum, the limb T-P profiles shown in Figure~\ref{t_p_fig} reveal a 1000 K temperature difference between the two terminators, which could further reduce the strength of spectral features on the cooler terminator by up to 50\%.  In contrast, high-altitude clouds will virtually erase spectral features. The aerosol effect is therefore predicted to be the dominant effect in differences detected between ingress and egress spectra. 

There are two primary observational challenges to obtaining high signal-to-noise ingress and egress spectra. (1) Only a fraction of the planet is in front of the star during these times, resulting in a transit depth that is on average only 50\% of the full-transit depth. (2) The ingress and egress durations are shorter than the transit duration by a factor of approximately $R_{pl}/R_*$ for a central transit. Still, for favorable transmission spectroscopy targets, the effects modeled in this Letter should ultimately be detectable in datasets that combine multiple transits and are timed to include the entirety of ingress and egress. For the example of WASP-121b, \citet{eva16}  obtained a high-fidelity, high signal-to-noise transmission spectrum with only 100 minutes of in-transit observations using the WFC3 instrument on \emph{HST}. Scaling their $\sim$100 ppm error bars to the 20 min ingress duration of WASP-121b requires that 5 transits must be observed with \emph{HST} to recover the same magnitude error bars on the ingress and egress spectra. This would place the differential aerosol signatures shown in Figure~\ref{diff_fig} within reach of detectability. Ultimately, with \emph{JWST} the aerosol effect will be observable with just a single transit of WASP-121b, and the characterization of additional less-favorable targets will be possible with a few transits each.

\acknowledgements
We thank the anonymous referees, Michael Line, and Emily Rauscher for helpful comments on this manuscript. E.M.-R.K. acknowledges support from the Research Corporation for Science Advancement through the Cottrell Scholar program and from Grinnell College's Harris Faculty Fellowship. J.L.B. acknowledges support from the David and Lucile Packard Foundation and NASA through STScI grants GO-14792 and 14793.  V.P. acknowledges support from the Sagan Postdoctoral Fellowship through the NASA Exoplanet Science Institute.


\end{document}